\documentclass[twocolumn,preprintnumbers,secnumarabic,amsmath,amssymb,superscriptaddress,nofootinbib,floatfix]{revtex4}

\usepackage{graphicx}% Include figure files
\usepackage{dcolumn}% Align table columns on decimal point
\usepackage{bm}% bold math

\usepackage{epsfig}
\usepackage{psfrag}
\usepackage{amsmath}
\usepackage{color}

\def\beq{\begin{equation}}
\def\eeq{\end{equation}}
\def\be{\begin{equation}}
\def\ee{\end{equation}}
\def\bea{\begin{eqnarray}}
\def\eea{\end{eqnarray}}

\def\<{\left\langle}
\def\>{\right\rangle}

\begin{document}

\preprint{NORDITA-2009-68}
\preprint{MPP-2009-174}

\medskip

\title{Quark mixing sum rules and the right unitarity triangle}

\author{Stefan Antusch} \email{antusch@mppmu.mpg.de}
\affiliation{Max-Planck-Institut f\"ur Physik (Werner-Heisenberg-Institut)
F\"ohringer Ring 6, D-80805 M\"unchen, Germany}
\author{Stephen F. King}\email{king@soton.ac.uk}
\affiliation{School of Physics and Astronomy, University of Southampton,
SO17 1BJ Southampton, United Kingdom}
\author{Michal Malinsk\'{y}}\email{malinsky@kth.se}
\affiliation{Department of Theoretical Physics, School of
Engineering Sciences, Royal Institute of Technology (KTH) --
AlbaNova University Center, Roslagstullsbacken 21, 106 91 Stockholm,
Sweden}
\author{Martin Spinrath} \email{spinrath@mppmu.mpg.de}
\affiliation{Max-Planck-Institut f\"ur Physik (Werner-Heisenberg-Institut)
F\"ohringer Ring 6, D-80805 M\"unchen, Germany}
%%%%%%%%%%%%%%%%%%%%%%%%%%%%%%%%%%

%%%%%%%%%%%%%%%%%%%%%%%%%%%%%%%%%%
\begin{abstract}
In analogy with the recently proposed lepton mixing sum rules,
we derive quark mixing sum rules for the case of hierarchical quark mass matrices with 1-3 texture zeros,
in which the separate up and down type 1-3 mixing angles are approximately zero,
and $V_{ub}$ is generated from $V_{cb}$ as a result of 1-2 up type quark mixing.
Using the sum rules, we discuss the phenomenological viability of such textures,
including up to four texture zeros,
and show how the right-angled unitarity triangle, i.e., $\alpha \approx 90^\circ$,
can be accounted for by a remarkably simple scheme involving
real mass matrices apart from a single element being purely imaginary.
In the framework of grand unified theories, we
show how the quark and lepton mixing sum rules may combine to yield an accurate
prediction for the reactor angle.
\end{abstract}

\maketitle

\section{Introduction}

The origin and nature of quark and lepton masses and mixings remains
one of the most intriguing questions left unanswered by the
standard model (SM) of particle physics. Within the SM, quark and lepton masses and mixings
arise from Yukawa couplings which are essentially free and undetermined.
In extensions such as Grand Unified Theories (GUTs), the Yukawa couplings
within a particular family may be related, but the mass hierarchy between
different families is not explained and supersymmetry (SUSY) does not shed any light
on this question either. Indeed, in the SM or GUTs, with or without SUSY,
a specific structure of the Yukawa matrices has no intrinsic meaning due to basis transformations
in flavour space. For example, one can always work in a basis in which, say, the up quark mass
matrix is taken to be diagonal with the quark sector mixing arising entirely from the
down quark mass matrix, or {\it vice versa}, and analogously in the lepton sector
(see e.g.\ \cite{King:2006hn}). This is symptomatic of the fact that neither the SM or
GUTs are candidates for a theory of flavour.

The situation changes somewhat once these theories are extended to include a family symmetry
spontaneously broken by extra Higgs fields called flavons. This approach has recently
received a massive impetus due to the discovery of neutrino mass and approximately
tri-bimaximal lepton mixing \cite{HPS} whose simple pattern strongly suggests some kind of a non-Abelian
discrete family symmetry might be at work, at least in the lepton sector, and, assuming a GUT-type of structure relating
quarks and leptons at a certain high energy scale, within the quark sector too. The observed neutrino flavour symmetry may arise
either directly or indirectly from a range of discrete symmetry groups \cite{King:2009ap}.
Examples of the direct approach, in which one or more generators of the discrete family symmetry
appears in the neutrino flavour group, are typically based on $S_4$ \cite{Lam:2009hn}
or a related group such as $A_{4}$ \cite{Ma:2007wu,Altarelli:2006kg}
or $PSL(2,7)$ \cite{King:2009mk}. Models of the indirect kind, in which
the neutrino flavour symmetry arises accidentally, include also $A_4$ \cite{King:2006np}
as well as $\Delta_{27}$ \cite{deMedeirosVarzielas:2006fc} and the continuous flavour symmetries like,
e.g., $SO(3)$ \cite{King:2006me} or $SU(3)$ \cite{King:2003rf} which  
accommodate the discrete groups above as subgroups \cite{deMedeirosVarzielas:2005qg}.

Theories of flavour based on a spontaneously broken family symmetry
are constructed in a particular basis in which the vacuum alignment
of the flavons is particularly simple. This then defines a preferred basis
for that particular model, which we shall refer to as the ``flavour basis.''
In such frameworks, the resulting low energy effective
Yukawa matrices are expected to have a correspondingly simple form
in the flavour basis associated with the high energy simple flavon
vacuum alignment.
This suggests that it may be useful to look for simple Yukawa matrix structures in a particular
basis, since such patterns may provide a bottom-up route towards a theory of flavour based on a spontaneously
broken family symmetry.

Unfortunately, experiment does not tell us directly the structure of the Yukawa matrices,
and the complexity of the problem, in particular, the basis ambiguity from the bottom-up perspective,
generally hinders the prospects of deducing even the basic features of the
underlying flavour theory from the experimental data. We are left with little alternative but to
follow an {\it ad hoc} approach pioneered some time ago by Fritzsch \cite{Fritzsch:1979zq,Fritzsch:1999ee} and currently
represented by the myriads of proposed
effective Yukawa textures (see e.g.\ \cite{Fritzsch:1979zq,Fritzsch:1999ee,textures,Roberts:2001zy,Chiu:2000gw,Fritzsch:1999rb})
whose starting assumption is that
(in some basis) the Yukawa matrices exhibit certain nice features such as symmetries or zeros
in specific elements which have become known as ``texture zeros.'' For example, in his classic paper,
Fritzsch pioneered the idea of having six texture zeros in the 1-1, 2-2, 1-3 entries of the Hermitian
up and down quark Yukawa (or mass) matrices \cite{Fritzsch:1979zq}.
Unfortunately, these six-zero textures are no longer consistent with experiment,
since they imply the bad prediction $|V_{cb}|\sim \sqrt{m_s /m_b}$,
so texture zerologists
have been forced to retreat to the (at most) four-zero schemes discussed,
for example, in \cite{Roberts:2001zy,Chiu:2000gw,Fritzsch:1999rb}
which give up on the 2-2 texture zeros allowing the good prediction $|V_{cb}|\sim m_s /m_b$.

However, four-zero textures featuring zeros in the 1-1 and 1-3  
entries of both up and down Hermitian mass matrices may also lead to the   
bad prediction $|V_{ub}|/|V_{cb}|\sim \sqrt{m_u /m_c}$
unless $|V_{cb}|$ results from the cancellation of quite sizeable up- and down-type quark 2-3 mixing
angles, leading to non-negligible induced 1-3 up- and down-type quark mixing \cite{Fritzsch:1999rb}.
Another possibility
is to give up on the 1-3 texture zeros, as well as the 2-2 texture zeros,
retaining only two texture zeros in the 1-1 entries of the up and down quark matrices \cite{Roberts:2001zy}.
Here we reject both of these options, and instead choose to
maintain up to four texture zeros, without invoking cancellations, for example by making
the 1-1 element of the up (but not down) quark mass matrix nonzero,
while retaining 1-3 texture zeros in both the up and down quark Hermitian matrices,
as suggested in \cite{Chiu:2000gw}.

In this paper we discuss phenomenologically viable textures 
for hierarchical quark mass matrices  
which have both
1-3 texture zeros and negligible 1-3 mixing in both the up and down quark mass matrices.
Such textures clearly differ from the textures discussed in \cite{Roberts:2001zy} and \cite{Fritzsch:1999rb},
but include some cases discussed in \cite{Chiu:2000gw}, as remarked above.
Our main contribution in this paper is to derive quark mixing sum rules
applicable to textures of this type, in which
$V_{ub}$ is generated from $V_{cb}$ as a result of 1-2 up-type mixing,
in direct analogy to the lepton sum rules derived in \cite{sumrule,Antusch:2008yc}.
Another important result of our study is to use the sum rules
to show how the right-angled unitarity triangle, i.e., $\alpha \approx 90^\circ$,
can be accounted for by a remarkably simple scheme involving
real mass matrices apart from a single element
of either the up or down quark mass matrix
being purely imaginary.
Fritzsch and Xing have previously emphasized how their four-zero scheme
with 1-1 and 1-3 texture zeros in the Hermitian up and down mass matrices
can be used to accommodate right unitarity triangles \cite{Fritzsch:1999rb},
but since their scheme involves large 2-3 and non-negligible 1-3 up and down quark mixing,
our sum rules are not applicable to their case.
Therefore, the textures in Refs.~\cite{Roberts:2001zy} and \cite{Fritzsch:1999rb} do not allow us to explain $\alpha \approx 90^\circ$ by simple structures 
with a combination of purely real and purely imaginary matrix elements. 
Recently, it has become increasingly clear that current data is indeed consistent with
the hypothesis of a right unitarity triangle, with the best fits giving
$\left( \alpha = 90.7^{+4.5}_{-2.9} \right)^\circ$
\cite{CKM}, and this provides additional impetus for our scheme.
The phenomenological observation that $\alpha \approx \pi /2$ has also motivated other approaches 
(see e.g. \cite{Xing, Scott, Toharia}) which are complementary to the approach developed in this paper.

The layout of the rest of the paper is as follows.
In Section 2, we derive the quark mixing sum rules, assuming zero up and down quark 1-3 mixing angles.
In Section 3, using the sum rules, we discuss the phenomenological viability of
quark mass matrix textures with 1-3 texture zeros, show how
modifications in the up sector can achieve viability,
and show how $\alpha \approx 90^\circ$ allows
each matrix element to be either real or purely imaginary. 
In Section 4,
in the framework of GUTs, we
discuss the implications of zero 1-3 mixing for the charged lepton and neutrino sectors,
and show that the quark mixing sum rules may be used to yield an accurate
prediction for the reactor angle.
Finally, Section 5 concludes the paper. Appendix A shows that textures with nonzero 1-3
elements in the up sector are disfavoured.

\section{Quark mixing sum rules from negligible 1-3 up and down mixing}

\subsection{Conventions}\label{sec:upanddown}
The mixing matrix in the quark sector, the Cabibbo-Kobayashi-Maskawa (CKM) matrix $U_{CKM}$,
is defined as the unitary matrix occurring in the charged current part
of the SM interaction Lagrangian expressed in terms of the quark
mass eigenstates. These mass eigenstates can be determined from the mass matrices
in the Yukawa sector, namely
\begin{equation}
\mathcal{L}_{Y}=-\overline{u^i_L} (M_u)_{ij} u^j_R - \overline{d^i_L}
(M_d)_{ij} d^j_R + H.c. \;,
\end{equation}
where $M_u$ and $M_d$ are the mass matrices of the up-type and
down-type quarks, respectively. The change from the flavour into
the mass basis is achieved via bi-unitary transformations
\begin{eqnarray}
V_{u_L} M_u V_{u_R}^\dagger &=& \mbox{diag}(m_u, m_c, m_t), \\
V_{d_L} M_d V_{d_R}^\dagger &=& \mbox{diag}(m_d, m_s, m_b),
\end{eqnarray}
where $V_{u_{L}}$, $V_{u_{R}}$, $V_{d_{L}}$ and $V_{d_{R}}$ are
unitary $3\times 3$ matrices. The CKM matrix $U'_{CKM}$ (in the ``raw'' form,
i.e.\ before the ``unphysical'' phases were absorbed into
redefinitions of the quark mass eigenstate field operators) is then given by
\begin{equation}\label{eq:UCKM_VuVd}
U'_{CKM} =V_{u_L} V_{d_L}^\dagger.
\end{equation}

In this paper we shall use the standard (or so-called Particle Data Group (PDG)
\cite{PDG}) parameterisation for the CKM matrix
(after eliminating the ``unphysical'' phases) with the structure
\be
\label{UPDG} U_{CKM}=R_{23}U_{13}R_{12} \;,
\ee
where
$R_{23},R_{12}$ denote real (i.e.\ orthogonal) matrices,
and the unitary matrix $U_{13}$ contains the observable phase $\delta_{\rm CKM}$.
For more details, see e.g.\ \cite{PDG}. Other alternative parametrisations, motivated by the observation of $\alpha \approx 90^\circ$ (see e.g.\ \cite{Xing:2009eg}) have been suggested,
but we prefer to stick to the standard one here.

However, in order to construct the ``physical'' CKM matrix $U_{CKM}$ in any given theory of flavour one should begin with the ``raw'' CKM matrix $U'_{CKM}$ defined in Eq.~(\ref{eq:UCKM_VuVd}), where $V_{u_L}$ and $V_{d_L}$ on the right-hand
side are general unitary matrices.
Recall that a generic 3$\times$3 unitary matrix $V^{\dagger}$ can be always written in
terms of three angles $\theta_{ij}$, three phases $\delta_{ij}$ (in all cases $i<j$) and three phases $\gamma_{i}$
in the form \cite{King:2002nf}
\be
\label{eq:param1}
V^{\dagger}=
U_{23} U_{13} U_{12} \,{\rm diag}(e^{i\gamma_1},e^{i\gamma_2},e^{i \gamma_3})\;,
\ee
where the
three unitary transformations $U_{23}, U_{13}, U_{12}$ are defined
as
\begin{eqnarray}\label{eq:U12}
U_{12}= \left(\begin{array}{ccc}
  c_{12} & s_{12}e^{-i\delta_{12}} & 0\\
  -s_{12}e^{i\delta_{12}}&c_{12} & 0\\
  0&0&1\end{array}\right)
\end{eqnarray}
(and analogously for $U_{13},U_{23}$). As usual, $s_{ij}$ and
$c_{ij}$ are abbreviations for $\sin \theta_{ij}$ and $\cos
\theta_{ij}$, and the $\theta_{ij}$ angles can be always made
positive by a suitable choice of the $\delta_{ij}$'s.
It is convenient to use this
parameterisation for both $V^\dagger_{u_L}$ and $V^\dagger_{d_L}$,
where the phases $\gamma_i$ can immediately be absorbed into the quark mass
eigenstates. Thus, they can be dropped and one is effectively left only with
\be
\label{eq:param2}
V^{\dagger}_{u_L}=
U^{u_L}_{23} U^{u_L}_{13} U^{u_L}_{12}\quad\mathrm{and}\quad
V^{\dagger}_{d_L}=U^{d_L}_{23} U^{d_L}_{13} U^{d_L}_{12},
\ee
where $V^{\dagger}_{u_L}$ involves the angles
$\theta^u_{ij}$ and phases $\delta^u_{ij}$,
while $V^{\dagger}_{d_L}$ involves the angles
$\theta^d_{ij}$ and phases $\delta^d_{ij}$.
Using Eqs.~(\ref{eq:UCKM_VuVd}) and (\ref{eq:param2}) $U'_{CKM}$ can be
written as
\be
\label{eq:param3}
U'_{CKM} = {U^{u_L}_{12}}^\dagger
{U^{u_L}_{13}}^\dagger {U^{u_L}_{23}}^\dagger
U^{d_L}_{23}U^{d_L}_{13} U^{d_L}_{12}
 \;.
\ee
On the other hand, $U'_{CKM}$ can be also parametrised along the lines of Eq.~(\ref{eq:param1}),
\be
\label{eq:param4}
U'_{CKM}=
U_{23} U_{13} U_{12} \, {\rm diag}(e^{i\gamma_1},e^{i\gamma_2},e^{i \gamma_3})\;.
\ee
By comparing Eq.~(\ref{eq:param4}) to Eq.~(\ref{UPDG}), we see that
the angles $\theta_{ij}$ are the standard PDG ones in $U_{CKM}$, and
five of the six phases of  $U'_{CKM}$ in Eq.~(\ref{eq:param4}) may be removed leaving the
standard PDG phase in $U_{CKM}$ identified as \cite{King:2002nf}
\be\label{eq:deltafromparam1}
\delta_{\rm CKM} = \delta_{13}-\delta_{23}-\delta_{12} \;.
\ee

\subsection{Mixing angle sum rules}
Let us now suppose that
$\theta_{13}^d = \theta_{13}^u =0$.
From the SM point of view,
this corresponds to just a convenient choice of basis but, as
discussed in the Introduction, it becomes a nontrivial assumption
at the level of a specific underlying model of flavour.
For models where zero 1-3 mixing is realised with flavour symmetries,
see e.g.\ \cite{Barr:1990td}.
For $\theta_{13}^d = \theta_{13}^u =0$, Eq.~(\ref{eq:param3}) simplifies to
\be
\label{eq:param5}
U'_{CKM} = {U^{u_L}_{12}}^\dagger
{U^{u_L}_{23}}^\dagger
U^{d_L}_{23} U^{d_L}_{12}
 \;.
\ee
Then, by equating the right-hand sides of Eqs.~(\ref{eq:param4}) and (\ref{eq:param5})
and expanding to leading order in the small mixing angles,
we obtain the following relations (up to cubic terms in the
physical quark mixing angles):
\bea \label{F1}
{\theta_{23}}e^{-i\delta_{23}}&=&
{\theta_{23}^{d}}e^{-i\delta_{23}^{d}}
-{\theta_{23}^{u}}e^{-i\delta_{23}^{u}}\;,
\\
\label{F2} {\theta_{13}}e^{-i\delta_{13}}&=&
-{\theta_{12}^{u}}e^{-i\delta_{12}^{u}}
({\theta_{23}^{d}}e^{-i\delta_{23}^{d}} - {\theta_{23}^{u}}e^{-i\delta_{23}^{u}})
\;,\\
\label{F3} {\theta_{12}}e^{-i\delta_{12}}&=&
{\theta_{12}^{d}}e^{-i\delta_{12}^{d}}
-{\theta_{12}^{u}}e^{-i\delta_{12}^{u}} \;.
\eea
Let us first consider Eq.~(\ref{F2}), which can be transformed into
\begin{eqnarray}\label{Eq:theta13withphases}
{\theta_{13}}e^{-i\delta_{13}}&=& -{\theta_{12}^{u}} {\theta_{23}}
e^{-i(\delta_{12}^{u}+\delta_{23})} \;,
\end{eqnarray}
where $\theta_{13}$ and $\theta_{23}$ stand for the measurable 1-3
and 2-3 mixing angles in the quark sector, respectively.  Taking the
modulus of  Eq.~(\ref{Eq:theta13withphases}), the 1-2 angle entering the up-sector rotation
($V_{u_L}$) in the flavour basis obeys
\begin{eqnarray}\label{Eq:QuarkRelation}
\label{Eq:theta13}{\theta_{12}^{u}} &=&
\frac{\theta_{13}}{\theta_{23}} = \left( 4.96 \pm 0.30
\right)^\circ \;.
\end{eqnarray}
where the 1$\sigma$ errors are displayed \cite{PDG}.

Similarly, combining Eq.~(\ref{F3}) with
Eq.~(\ref{Eq:theta13withphases}) one receives
\begin{eqnarray}
{\theta_{12}} - \frac{\theta_{13}}{\theta_{23}} e^{-i(\delta_{13}
- \delta_{23} - \delta_{12})} &=&
{\theta_{12}^{d}}e^{-i(\delta_{12}^{d} -  \delta_{12})}\;.
\end{eqnarray}
This, together with the identification
Eq.~(\ref{eq:deltafromparam1}) gives rise to the quark
sector sum rule\footnote{ We would like to remark that for
$\theta_{12}^{u} \ll \theta_{12}^{d}$, the sum rule may be further
simplified to $\theta_{12} - \frac{\theta_{13}}{\theta_{23}} \cos
\delta = \theta_{12}^d$. For similar considerations in the lepton
sector, see e.g.\ \cite{sumrule}.}
\begin{eqnarray}\label{Eq:QuarkSumRule}
\theta_{12}^d = \left|\theta_{12} - \frac{\theta_{13}}{\theta_{23}} e^{-
i\delta_\mathrm{CKM}} \right| = \left(
12.0^{+0.39}_{-0.22} \right)^\circ  
\end{eqnarray}
which is valid up to higher order corrections. The present best-fit value and the 1$\sigma$ errors are also displayed.

Needless to say, the relations (\ref{Eq:QuarkRelation}) and
(\ref{Eq:QuarkSumRule}) apply at the scale at which the flavour structure
emerges, often close to the scale of Grand Unification.
Thus, in principle, the renormalisation group (RG) effects should
be taken into account. However, due to the smallness of the mixing
angles in the quark sector and the hierarchy of the quark masses,
the RG corrections to the above relations are very small and can be neglected to a very good
approximation.

\subsection{Phase sum rule}

It is interesting that, with the 1-2 mixing angles in the up and down
sector derived from the physical parameters,
the 1-2 phase difference in the up and down sectors can
also be determined. Indeed, combining all three equations
(\ref{F1})-(\ref{F3}), one obtains
\begin{eqnarray} \label{Eq:deltadu}
\frac{\theta_{13} \theta_{12}}{\theta_{23}} e^{i \delta_{\rm CKM}}
&=& - \theta_{12}^{u}
({\theta_{12}^{d}}e^{-i(\delta_{12}^{d}-\delta_{12}^{u})}-{\theta_{12}^{u}}) \;.
\end{eqnarray}
Using Eqs.~(\ref{Eq:QuarkRelation}) and (\ref{Eq:QuarkSumRule}) we can solve Eq.~(\ref{Eq:deltadu})
for $\delta_{12}^{d} - \delta_{12}^{u}$ and obtain (at 1$\sigma$ level)
\begin{eqnarray}
\delta_{12}^{d} - \delta_{12}^{u} = (91.5^{+5.5}_{-4.0})^\circ\;,
\label{phasediff}
\end{eqnarray}
which is remarkably close to $\pi/2$.
We emphasise that this is a consequence of zero 1-3
mixing in the up and down sectors, $\theta_{13}^d = \theta_{13}^u = 0$.

We now show that, assuming quark textures with negligible 1-3 up and down quark mixing,
corresponding to 1-3 texture zeros for hierarchical quark mass matrices, 
$\delta_{12}^{d} - \delta_{12}^{u}$
is approximately equal to $\alpha$.
This comes from the definition of the unitarity triangle angle $\alpha$:
\begin{eqnarray}
 \alpha &=& \arg \left( - \frac{V_{td} V_{tb}^*}{V_{ud} V_{ub}^*} \right) \nonumber \\
& = & \arg \left( - \frac{(s_{12} s_{23} - c_{12} c_{23} s_{13} e^{i \delta})
c_{23} c_{13}}{ c_{12} c_{13} s_{13} e^{i\delta}} \right)  \nonumber \\
& \approx & \arg \left( 1 -
\frac{\theta_{12} \theta_{23} }{\theta_{13}} e^{-i\delta} \right)\,.
\end{eqnarray}
For the second term in the argument, we can use Eqs.~(\ref{F1}), (\ref{F2}) and (\ref{F3})
\begin{eqnarray}
 \alpha &\approx& \arg \left( 1 +\frac{\theta_{23} e^{i\delta_{23} }
 (\theta_{12}^d e^{i\delta_{12}^d} - \theta_{12}^u e^{i \delta_{12}^u})}{\theta_{12}^u e^{i \delta_{12}^u} \theta_{23}
 e^{i\delta_{23} } } \right) \nonumber\\
&=& \arg \left( \frac{\theta_{12}^d}{\theta_{12}^u} e^{i
(\delta_{12}^d - \delta_{12}^u)} \right) = \delta_{12}^d -
\delta_{12}^u \;.
\end{eqnarray}
Thus, one can see that the angle $\alpha$ is nothing but the phase
difference $\delta_{12}^d - \delta_{12}^u$, corresponding to a very simple phase sum rule
\be
\label{phase}
 \alpha \approx \delta_{12}^d - \delta_{12}^u \;.
\ee

\section{Quark mass matrices with 1-3 texture zeros}

\subsection{Real/imaginary matrix elements for $\alpha = 90^\circ$}\label{Sec:deltaCKM}
According to the phase sum rule in Eq.~(\ref{phase}),
the experimental observation that
$\alpha \approx 90^\circ$,
or the equivalent determination
in Eq.~(\ref{phasediff}), suggests looking at quark mass matrices with 1-3 texture zeros and with
$\delta_{12}^{d}$ or $\delta_{12}^{u}$ at the special values $\pm \pi/2$.
This would correspond to a set of rather specific textures
of the quark mass matrices with, for example, purely imaginary 1-2
elements in either $M_u$ or $M_d$ while the 2-2 elements remain
real. For example, in \cite{King:2002nf} the relation between the phases of the mixing angles and the phases of the matrix elements is discussed.  For instance, the following patterns naturally emerge:
\begin{eqnarray}
M_u =
\begin{pmatrix}
a_u   & -i b_u & 0 \\
* & c_u & d_u \\
*   & *   & e_u
\end{pmatrix}
\; , \quad
M_d =
\begin{pmatrix}
a_u   &  b_d & 0 \\
* & c_d & d_d \\
*   & *   & e_d
\end{pmatrix}
\end{eqnarray}
and/or
\begin{eqnarray}
M_u =
\begin{pmatrix}
a_u   &  b_u & 0 \\
* & c_u & d_u \\
*   & *   & e_u
\end{pmatrix}
\; , \quad
M_d =
\begin{pmatrix}
a_u   & i b_d & 0 \\
* & c_d & d_d \\
*   & *   & e_d
\end{pmatrix} ,
\end{eqnarray}
where $a_u,b_u,c_u,d_u,e_u$ and $a_d,b_d,c_d,d_d,e_d$ are real
parameters, and the elements marked by ``*'' are irrelevant as long
as the hierarchy of the mass matrix is large enough, or,
equivalently, as long as the mixing angles in $V_{u_R}$ and
$V_{d_R}$ are small. These textures are all phenomenologically viable,
and consistent with $\alpha= 90^\circ$, and their simple phase structure
provides a post justification of our assumption of 1-3 texture zeros
and negligible 1-3 up- and down-type quark mixing.
However, the above textures are clearly not the most predictive ones and, for example,
do not relate the up and down quark 1-2 mixing angles to
masses. This requires additional assumptions,
such as additional texture zeros and Hermitian or symmetric matrices, as we now discuss.

\subsection{Four-zero textures confront the sum rules }\label{Sec:SumrulePlusGST}
Under the additional assumptions of symmetric or Hermitian mass matrices in the
1-2 block and zero textures in the 1-1 positions of the quark mass
matrices, i.e.,
\begin{eqnarray}\label{Eq:texture0}
M_u =
\begin{pmatrix}
0   & b_u & 0 \\
b_u & c_u & d_u \\
*   & *   & e_u
\end{pmatrix}
\; , \quad M_d =
\begin{pmatrix}
0   & i b_d & 0 \\
\pm i b_d & c_d & d_d \\
*   & *   & e_d
\end{pmatrix}
\end{eqnarray}
we obtain as additional predictions the  Gatto-Sartori-Tonin (GST) relations \cite{GST}
with 1$\sigma$ errors displayed,
\begin{eqnarray}
\label{eq:m_u_over_m_c}\theta_{12}^{u} &=& \sqrt{\frac{m_u}{m_c}} = \left( 2.61^{+0.54}_{-0.46} \right)^\circ \;, \\
\label{eq:m_d_over_m_s}\theta_{12}^{d} &=& \sqrt{\frac{m_d}{m_s}} = \left( 13.2^{+3.4}_{-3.3} \right)^\circ\;.
\end{eqnarray}
Here we already see a conflict in the up sector.
The prediction for $\theta_{12}^u$ from the sum rule in Eq.\ (\ref{Eq:QuarkRelation}) is quite different
(several $\sigma$ away) from the GST relation above.
That suggests that the texture in the up sector should be modified to be in good agreement with experiment.
By contrast the prediction from the sum rule in Eq.\ (\ref{Eq:QuarkSumRule}) for $\theta_{12}^d$ is in very good agreement
(within the errors) with the GST result in Eq.\ (\ref{eq:m_d_over_m_s}), and therefore it is quite plausible to
keep the simple texture ansatz for the down sector.

Combining Eqs.\ (\ref{eq:m_u_over_m_c}) and (\ref{eq:m_d_over_m_s}) with the sum rules in
Eqs.~(\ref{Eq:QuarkRelation}) and (\ref{Eq:QuarkSumRule}), the two relations
\be
\label{Eq:QSplusGST1}
\left|\theta_{12} -
\frac{\theta_{13}}{\theta_{23}} e^{- i\delta_\mathrm{CKM}} \right|
= \sqrt{\frac{m_d}{m_s}}
\ee
and
\be
\label{Eq:QSplusGST2}
\frac{\theta_{13}}{\theta_{23}}=\sqrt{\frac{m_u}{m_c}}
\ee
emerge.
We emphasize that these results do not hold for the textures in \cite{Fritzsch:1999rb}
where the 2-3 up and down quark mixings are large and the 1-3 up and down quark mixings are
non-negligible.

\begin{figure}
 \centering
 \includegraphics[scale=0.485]{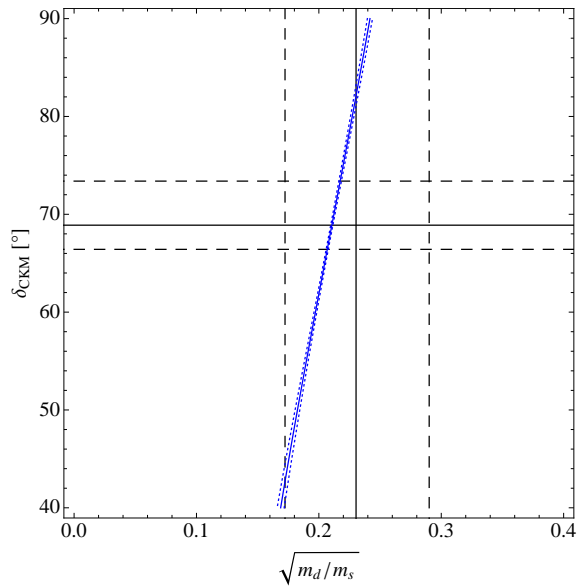}
 \caption{Graphical illustration of the relation of Eq.~(\ref{Eq:QSplusGST1}).
 The blue lines indicate the predicted values of
 $\delta_\mathrm{CKM}$ for given $\sqrt{m_d/m_s}$
 under the assumptions of section \ref{Sec:SumrulePlusGST},
 and the dashed horizontal and vertical black lines (and solid black lines)
 show the $1\sigma$ errors (and best-fit values) for
 $\delta_\mathrm{CKM}$ and $\sqrt{m_d/m_s}$, respectively. }
\end{figure}

The compatibility of Eq.\ (\ref{Eq:QSplusGST1}) with
the experimental results for the down-type quark masses and mixing
parameters \cite{PDG} is illustrated in Fig.\ 1. 
We note that RG running for the quark masses, as well as their
potential SUSY threshold corrections, are very similar for the
first two generations and thus cancel out in their ratio. For our
estimates, we have considered the running quark masses at the top
mass scale $m_t (m_t)$ \cite{Xing:2007fb}.
$\delta_\mathrm{CKM}$ is extracted for given $\sqrt{m_d/m_s}$. The
solid blue line shows the relation for best-fit values of the
parameters while the dashed blue lines indicate the range with
$1\sigma$ errors included. The dashed horizontal and vertical
black lines (and solid black lines) show the $1\sigma$ errors (and
best-fit values) for $\delta_\mathrm{CKM}$ and $\sqrt{m_d/m_s}$,
respectively. The relation of Eq.\ (\ref{Eq:QSplusGST1}) is well
compatible with the present data.
Future more precise experimental
measurements (for instance at LHCb or $B$ factories) and, in particular,
an improved knowledge on $m_d$ (e.g.\ from lattice QCD) are
required to test it more accurately.

In the following, we consider some examples of possible
modifications to the textures in the up sector which are
phenomenologically acceptable, while leaving the successful down sector texture unchanged,
and retaining the successful real and imaginary scheme which leads to the right unitarity triangle.
As discussed in Appendix A,
the idea of relaxing the up quark 1-3 texture zero is disfavoured,
so we restrict ourselves to either relaxing the up quark 1-1 texture zero,
or relaxing symmetry in the 1-2 up quark sector, as discussed below.

\subsection{Relaxing the up quark 1-1 texture zero}
One possible modification is to introduce a nonzero element in
the 1-1 position of the up quark mass matrix, i.e.\
\begin{eqnarray}
M_u =
\begin{pmatrix}
a_u   & b_u & 0 \\
b_u & c_u & d_u \\
*   & *   & e_u
\end{pmatrix}
\; , \quad M_d =
\begin{pmatrix}
0   & i b_d & 0 \\
\pm i b_d & c_d & d_d \\
*   & *   & e_d
\end{pmatrix}.
\end{eqnarray}
As a result, we obtain the up sector relation
\begin{eqnarray}
m_u = a_u - \frac{b^2_u}{c_u}
\end{eqnarray}
which allows to adjust $a_u$, which is of the order of the up quark mass, while $b_u/c_u \approx
\theta_{12}^u$ has to be equal to the value obtained in
Eq.~(\ref{Eq:QuarkRelation}) using the sum rule. For the
down sector, there is still the successful prediction from Eq.\ (\ref{eq:m_d_over_m_s}) 
leading to the successful sum rule relation of
Eq.~(\ref{Eq:QSplusGST1}). Furthermore, as discussed in section
\ref{Sec:deltaCKM}, the Dirac phase of the CKM matrix is correct. We note that there exist several variants of the
texture.
For example, we can choose the 1-2 element of $M_d$ real,
the 1-2 element of $M_u$ purely imaginary and all the other elements also
real.
These variants are valid as long as $b_u/c_u$
is real and $b_d/c_d$ is purely imaginary, or {\em vice versa}.

We emphasize that the elements marked by ``*'' are irrelevant as
long as the hierarchy of the mass matrix is large enough, so they
may be replaced by zeros or, if the matrices are Hermitian, the
3-1 elements may be zero while the 3-2 elements are determined by
Hermiticity. Since the sum rule in Eq.\ (\ref{F1}) shows that $V_{cb}$
is determined only by the difference in 2-3 mixing angles in the
up and down sectors, it is also possible to set either $d_u$
or $d_d$ equal to zero without changing the physical predictions.
In this way it is possible to arrive at some of the four-zero textures
discussed, for example, in \cite{Chiu:2000gw}. However, we emphasize that
here we are additionally assuming the real and imaginary structures consistent
with the right unitarity triangle and this was not discussed in \cite{Chiu:2000gw}.

\subsection{Relaxing the up quark 1-2 symmetry}
A second option for a texture consistent with experimental data consists in relaxing the symmetry of the 1-2 block in the up sector, while keeping the texture zero in the 1-1 position:
\begin{eqnarray}
M_u =
\begin{pmatrix}
0   & b_u & 0 \\
b'_u & c_u & d_u \\
*   & *   & e_u
\end{pmatrix}
\; , \quad
M_d =
\begin{pmatrix}
0   & i b_d & 0 \\
\pm i b_d & c_d & d_d \\
*   & *   & e_d
\end{pmatrix}.
\end{eqnarray}
The two up-sector relations
\begin{equation}
m_u \approx b_u b'_u/c_u\;,\quad
m_c \approx c_u\;,\quad
\theta_{12}^u = \frac{b_u}{c_u}
\end{equation}
can be simultaneously fulfilled by choosing $b_u,b'_u$ and $c_u$ appropriately.
The prediction from Eq.\ (\ref{eq:m_d_over_m_s}) 
and the prediction for $\delta_{\rm CKM}$ do not change and remain compatible with data. We note that there exist several variants of the texture. As before, it is sufficient to have $b_u/c_u$ real and $b_d/c_d$  purely imaginary, or {\em vice versa}.

\section{Quark-lepton mixing relations}
Extending the notion of zero 1-3 mixing to the lepton sector (i.e.\ under the assumption of
$\theta^\nu_{13} = \theta^e_{13} = 0$), the presently unknown
mixing angle $\theta^{MNS}_{13}$ of the leptonic (MNS) mixing
matrix satisfies the relation (analogous to Eq.~(\ref{Eq:QuarkRelation}))
\be
\theta^{MNS}_{13} = \sin \theta^{MNS}_{23} \theta^e_{12} \;,
\ee
where $\theta^e_{12}$ is
the 1-2 mixing in the charged lepton mass matrix $M_e$.
This relation has emerged before, for example, in
the context of lepton sum rules in \cite{sumrule}.

In many classes of GUT models of flavour, the 1-2 mixing angles corresponding to $M_e$
and $M_d$ are related by a group theoretical Clebsch factor, for
example $\theta^d_{12} = 3 \theta^e_{12}$ \cite{GJ}.
In general, it is usually assumed that $\theta^d_{12}$ is of the order of the
Cabibbo angle, leading to a prediction $\theta^{MNS}_{13}\sim 3^\circ$ \cite{sumrule}.
However, in the context of Fritzsch-type textures, which are based on Hermitian matrices with 1-1, 2-2 and 1-3 texture zeros, this prediction can be made
more precise by using the sum rule which relates $\theta^d_{12}$ to down-type quark masses.
Thus, applying Eq.~(\ref{Eq:QuarkSumRule}) at low energies and taking the present
experimental data for the quark mixing angles, and for
$\theta^{MNS}_{23}$ (taken from \cite{Schwetz:2008er}), one can
make the rather precise prediction
\be
\label{Eq:MNS13mixing}
\theta^{MNS}_{13} = \left( 2.84^{+0.22}_{-0.18} \right)^\circ
\ee
which gives $\sin^2 \theta^{MNS}_{13} =
0.0025^{+0.0004}_{-0.0003}$ and holds under the assumption of
texture zeros in the 1-3 elements of the mass matrices (or more
precisely $\theta^u_{13} =\theta^d_{13} =\theta^\nu_{13} =
\theta^e_{13} = 0$) and $\theta^d_{12} = 3 \theta^e_{12}$. Of
course, Eq.~(\ref{Eq:MNS13mixing}) is only a single example out of a
larger variety of predictions which may arise in unified flavour
models (see e.g.\ \cite{Antusch:2009gu}).
We emphasise that the main use of Eq.~(\ref{Eq:QuarkSumRule}) in this context is
that it allows to ``determine'' the down quark mixing
$\theta^d_{12}$, which is generically involved in relations
between quark and lepton mixing angles, from measurable
quantities.

We note that in the lepton sector, the RG
corrections (see e.g.\ \cite{RGE}) can be significant, depending on the absolute neutrino mass
scale (and on $\tan \beta$ in a SUSY framework) and
other effects such as canonical normalisation on the mixing
angles can also be sizeable \cite{Antusch:2008yc}.
Furthermore, relaxing the 1-1 texture zero in the up quark sector may switch on a nonzero
1-3 mixing angle in the neutrino sector via partially constrained sequential dominance
 \cite{King:2009qt}.

\section{Conclusions}

In this paper we have discussed phenomenologically viable textures 
for hierarchical quark mass matrices 
which have both
1-3 texture zeros and negligible 1-3 mixing in both the up and down quark mass matrices.
Such textures differ from the textures discussed in \cite{Roberts:2001zy} and \cite{Fritzsch:1999rb}.
Our main contribution in this paper has been to derive quark mixing sum rules
applicable to textures of this type, in which
$V_{ub}$ is generated from $V_{cb}$ as a result of 1-2 up-type mixing,
in direct analogy to the lepton sum rules derived in \cite{sumrule,Antusch:2008yc}.
An important result of our study is to show how the right-angled unitarity triangle, i.e., $\alpha \approx 90^\circ$,
can be accounted for by a remarkably simple scheme involving
real mass matrices apart from a single element 
of either the up or down quark mass matrix
being purely imaginary. The experimental result that $\alpha \approx 90^\circ$ therefore provides
an impetus for having hierarchical textures compatible with
negligible 1-3 mixing in both the up and down quark mass matrices.
This is probably the most important take-home message of this paper.

The quark mixing sum rules in Eqs.~(\ref{Eq:QuarkRelation}) and (\ref{Eq:QuarkSumRule})
relate the up and down quark 1-2 mixing angles to observable
parameters in the CKM matrix.
Using these sum rules the four-zero texture with 1-1 and 1-3 texture zeros
and a 2-1 symmetric or Hermitian structure,
is shown to be viable for the down quark sector but not for the
up quark sector. However, it is possible to have four-zero textures
compatible with our sum rules by, for example, filling in the up quark
1-1 texture zero, then having Hermitian matrices with either of the 2-3 elements in the
up or down sector set equal to zero as in \cite{Chiu:2000gw}. However, we emphasize that
here we are additionally assuming the real and imaginary structures consistent
with the right unitarity triangle and this was not discussed in \cite{Chiu:2000gw}.

In the framework of GUTs, it is natural
to have 1-3 texture zeros for both the quark and charged lepton sectors,
and in such a case we have shown that the quark mixing sum rules may be used to yield an accurate
prediction for the reactor angle; see Eq.~(\ref{Eq:MNS13mixing}). However, we caution that this prediction
is subject to considerable theoretical uncertainty due to the model dependence of the quark-lepton mixing angle
relations, RG and canonical normalisation effects, as well as the assumption that the
underlying neutrino 1-3 mixing angle is zero.
Indeed, relaxing the 1-1 texture zero in the up quark sector will switch on a nonzero
1-3 mixing angle in the neutrino sector via partially constrained sequential dominance
 \cite{King:2009qt}.

Finally, we emphasise that the strategy of exploring particular
textures for Yukawa matrices, though necessarily rather {\it ad hoc},
is meaningful from the perspective of theories of flavour based on
spontaneously broken family symmetry, where, in the flavour basis defined
by the high energy theory, simple Yukawa matrix structures are expected.
Indeed, the study of simple Yukawa textures may provide the only bottom-up
way of deducing a high energy theory of flavour from experimental data.
We have shown that $\alpha \approx 90^\circ$ may provide a clue
towards such a high energy theory of flavour
via rather simple Yukawa matrices involving 1-3 texture zeros whose nonzero
elements are either real or purely imaginary.
Such patterns could be achieved, in principle, by
appropriate alignment of the vacuum expectation values of flavour
symmetry breaking flavon fields, and in particular their phases. It would be interesting to build a theory
of flavour along these lines.

\section*{Acknowledgments}
We would like to thank Jonathan Flynn for discussions and Christoph Luhn 
for carefully reading the manuscript and providing helpful comments.
We are indebted to
NORDITA for the hospitality and support during the programme
``Astroparticle Physics -- A Pathfinder to New Physics'' held in
Stockholm in March 30 - April 30, 2009 during which part of this
study was performed. 
SFK is very grateful for the support and hospitality provided by the 
Max-Planck-Institute of Physics 
during his stay in Munich while this paper was being finalized.
SA and MS acknowledge partial support by the DFG cluster of excellence
``Origin and Structure of the Universe.''
The work of MM is supported by the Royal
Institute of Technology (KTH), Contract No. SII-56510.
SFK acknowledges partial support from the STFC
Rolling Grant No. ST/G000557/1 and a Royal Society Leverhulme Trust Senior Research Fellowship.

\begin{appendix}
\section{Textures with nonzero 1-3 elements are disfavoured}
With nonzero 1-3 elements, $\delta_{\rm CKM}$ depends not only on $\delta_{12}^{d} -
\delta_{12}^{u}$ but also on other parameters (in
particular $\delta_{13}^{u,d}$ and $\delta_{23}^{u,d}$) and the simple quark mixing sum rules in
Eqs.~(\ref{Eq:QuarkRelation}) and (\ref{Eq:QuarkSumRule}) are no longer valid.
Examples of this type of texture include (with real parameter $f_u$)
\begin{eqnarray} \label{Eq:fuTexture1}
M_u =
\begin{pmatrix}
0   & b_u & f_u \\
b_u & c_u & d_u \\
*   & *   & e_u
\end{pmatrix}
\; , \quad
M_d =
\begin{pmatrix}
0   & i b_d & 0 \\
i b_d & c_d & d_d \\
*   & *   & e_d
\end{pmatrix}
\end{eqnarray}
and
\begin{eqnarray}
M_u =
\begin{pmatrix}
0   &  b_u & i f_u \\
 b_u & c_u & d_u \\
*   & *   & e_u
\end{pmatrix}
\; , \quad
M_d =
\begin{pmatrix}
0   &  i b_d & 0 \\
i b_d & c_d & d_d \\
*   & *   & e_d
\end{pmatrix}
\end{eqnarray}
but also variations with different elements chosen either purely imaginary or real.

We will demonstrate our approach for this case by means of the texture in Eq.~(\ref{Eq:fuTexture1}). The starting point is here (similar to Eqs.~(\ref{F1})-(\ref{F3}))
\bea \label{G1}
{\theta_{23}}e^{-i\delta_{23}}&=&
{\theta_{23}^{d}}e^{-i\delta_{23}^{d}}
-{\theta_{23}^{u}}e^{-i\delta_{23}^{u}}\;,
\\
\label{G2} {\theta_{13}}e^{-i\delta_{13}}&=&
-{\theta_{13}^{u}}e^{-i\delta_{13}^{u}}
-{\theta_{12}^{u} \theta_{23}}e^{-i(\delta_{12}^{u} + \delta_{23})} 
\;,\\
\label{G3} {\theta_{12}}e^{-i\delta_{12}}&=&
{\theta_{12}^{d}}e^{-i\delta_{12}^{d}}
-{\theta_{12}^{u}}e^{-i\delta_{12}^{u}} \;,
\eea
where we have also neglected terms of order ${\cal O}(\theta_{13} \theta_{ij})$. From our texture ansatz, we know the phases $\delta_{12}^{u/d}$, $\delta_{23}^{u/d}$ and $\delta_{13}^{u}$. For the values of $\theta_{12}^{u/d}$, we take the values from the GST relations, i.e.\ Eq.~(\ref{eq:m_u_over_m_c}),
which hold here because of the zeros in the 1-1 position and the symmetric structure for the first two generations.  Then we can calculate $\theta_{13}^{u}$ and $\delta_{\rm CKM}$ in terms of the known quantities and obtain $\delta_{\rm CKM} = (78.83^{+3.62}_{-3.35})^\circ$. This result is several standard deviations away from the measurements.

Beyond the particular example discussed above, we found that the
inconsistency of the prediction for $\delta_{\rm CKM}$ also
appears in all other cases with $\delta_{23}$, $\delta_{12}^{u/d}$
and $\delta_{13}^{u/d}$ $\in \{0,\pm \pi/2\}$. Furthermore, the same happens for textures with $f_u = 0$ and $f_d \neq 0$, where $f_d$
denotes the 1-3 element of $M_d$. We conclude that under these conditions
textures with nonzero 1-3 elements are disfavoured.

\end{appendix}


\begin{thebibliography}{00}

%\cite{King:2006hn}
\bibitem{King:2006hn}
  S.~F.~King,
  %``Invariant see-saw models and sequential dominance,''
  Nucl.\ Phys.\  B {\bf 786} (2007) 52
  [arXiv:hep-ph/0610239].
  %%CITATION = NUPHA,B786,52;%%

\bibitem{HPS}
%\cite{Harrison:2002er}
%bibitem{Harrison:2002er}
P.~F.~Harrison, D.~H.~Perkins and W.~G.~Scott,
%``Tri-bimaximal mixing and the neutrino oscillation data,''
Phys.\ Lett.\ B {\bf 530} (2002) 167 [arXiv:hep-ph/0202074].
%%CITATION = HEP-PH 0202074;%%




%\cite{King:2009ap}
\bibitem{King:2009ap}
  S.~F.~King and C.~Luhn,
  %``On the origin of neutrino flavour symmetry,''
  arXiv:0908.1897 [hep-ph].
  %%CITATION = ARXIV:0908.1897;%%

%\cite{Lam:2009hn}
\bibitem{Lam:2009hn}
  C.~S.~Lam,
  %``A bottom-up analysis of horizontal symmetry,''
  arXiv:0907.2206 [hep-ph].
  %%CITATION = ARXIV:0907.2206;%%

%\cite{Ma:2007wu}
\bibitem{Ma:2007wu}
  E.~Ma and G.~Rajasekaran,
  %``Softly broken A(4) symmetry for nearly degenerate neutrino masses,''
  Phys.\ Rev.\  D {\bf 64} (2001) 113012
  [arXiv:hep-ph/0106291].
  %%CITATION = PHRVA,D64,113012;%%

%\cite{Altarelli:2006kg}
\bibitem{Altarelli:2006kg}
 %\cite{Altarelli:2006ri}
%\bibitem{Altarelli:2006ri}
  G.~Altarelli,
  %``Models of neutrino masses and mixings,''
  arXiv:hep-ph/0611117;
  %%CITATION = HEP-PH/0611117;%%
G.~Altarelli, F.~Feruglio and Y.~Lin,
  %``Tri-bimaximal neutrino mixing from orbifolding,''
  Nucl.\ Phys.\  B {\bf 775} (2007) 31
  [arXiv:hep-ph/0610165];
  %%CITATION = NUPHA,B775,31;%%
%\cite{Altarelli:2005yx}
%\bibitem{Altarelli:2005yx}
  G.~Altarelli and F.~Feruglio,
  %``Tri-bimaximal neutrino mixing, A(4) and the modular symmetry,''
  Nucl.\ Phys.\  B {\bf 741} (2006) 215
  [arXiv:hep-ph/0512103];
  %%CITATION = NUPHA,B741,215;%%
%\cite{Altarelli:2005yp}
%\bibitem{Altarelli:2005yp}
  G.~Altarelli and F.~Feruglio,
  %``Tri-bimaximal neutrino mixing from discrete symmetry in extra
  %dimensions,''
  Nucl.\ Phys.\  B {\bf 720} (2005) 64
  [arXiv:hep-ph/0504165];
  %%CITATION = NUPHA,B720,64;%%
%\cite{Chen:2009um}
%\bibitem{Chen:2009um}
  M.~C.~Chen and S.~F.~King,
  %``A4 See-Saw Models and Form Dominance,''
  JHEP {\bf 0906} (2009) 072
  [arXiv:0903.0125 [hep-ph]];
  %%CITATION = JHEPA,0906,072;%%
%\cite{Burrows:2009pi}
%\bibitem{Burrows:2009pi}
  T.~J.~Burrows and S.~F.~King,
  %``$A_4$ Family Symmetry from SU(5) SUSY GUTs in 6d,''
  arXiv:0909.1433 [hep-ph].
  %%CITATION = ARXIV:0909.1433;%%


%\cite{King:2009mk}
\bibitem{King:2009mk}
  S.~F.~King and C.~Luhn,
  %``A new family symmetry for SO(10) GUTs,''
  Nucl.\ Phys.\  B {\bf 820} (2009) 269
  [arXiv:0905.1686 [hep-ph]].
  %%CITATION = NUPHA,B820,269;%%

%\cite{King:2006np}
\bibitem{King:2006np}
  S.~F.~King and M.~Malinsky,
  %``A4 family symmetry and quark-lepton unification,''
  Phys.\ Lett.\  B {\bf 645} (2007) 351
  [arXiv:hep-ph/0610250].
  %%CITATION = PHLTA,B645,351;%%


%\cite{deMedeirosVarzielas:2006fc}
\bibitem{deMedeirosVarzielas:2006fc}
  I.~de Medeiros Varzielas, S.~F.~King and G.~G.~Ross,
  %``Neutrino tri-bi-maximal mixing from a non-Abelian discrete family
  %symmetry,''
  Phys.\ Lett.\  B {\bf 648} (2007) 201
  [arXiv:hep-ph/0607045].
  %%CITATION = PHLTA,B648,201;%%

%\cite{King:2006me}
\bibitem{King:2006me}
  S.~F.~King and M.~Malinsky,
  %``Towards a complete theory of fermion masses and mixings with SO(3)  family
  %symmetry and 5d SO(10) unification,''
  JHEP {\bf 0611} (2006) 071
  [arXiv:hep-ph/0608021];
  %%CITATION = JHEPA,0611,071;%%
S.~F.~King,
%``Predicting neutrino parameters from SO(3) family symmetry and quark-lepton
%unification,''
JHEP {\bf 0508} (2005) 105;
%[arXiv:hep-ph/0506297].
S.~Antusch and S.~F.~King,
  %``From hierarchical to partially degenerate neutrinos via type II upgrade  of
  %type I see-saw models,''
  Nucl.\ Phys.\  B {\bf 705} (2005) 239
  [arXiv:hep-ph/0402121].
  %%CITATION = NUPHA,B705,239;%%


%\cite{King:2003rf}
\bibitem{King:2003rf}
%\bibitem{King:2001uz}
  S.~F.~King and G.~G.~Ross,
  %``Fermion masses and mixing angles from SU(3) family symmetry,''
  Phys.\ Lett.\  B {\bf 520} (2001) 243
  [arXiv:hep-ph/0108112];
  %%CITATION = PHLTA,B520,243;%%
  S.~F.~King and G.~G.~Ross,
  %``Fermion masses and mixing angles from SU(3) family symmetry and
  %unification,''
  Phys.\ Lett.\  B {\bf 574} (2003) 239
  [arXiv:hep-ph/0307190];
  %%CITATION = PHLTA,B574,239;%%
%\cite{deMedeirosVarzielas:2005ax}
%\bibitem{deMedeirosVarzielas:2005ax}
  I.~de Medeiros Varzielas and G.~G.~Ross,
  %``SU(3) family symmetry and neutrino bi-tri-maximal mixing,''
  Nucl.\ Phys.\  B {\bf 733} (2006) 31
  [arXiv:hep-ph/0507176].
  %%CITATION = NUPHA,B733,31;%%


%\cite{deMedeirosVarzielas:2005qg}
\bibitem{deMedeirosVarzielas:2005qg}
  I.~de Medeiros Varzielas, S.~F.~King and G.~G.~Ross,
  %``Tri-bimaximal neutrino mixing from discrete subgroups of SU(3) and  SO(3)
  %family symmetry,''
  Phys.\ Lett.\  B {\bf 644} (2007) 153
  [arXiv:hep-ph/0512313].
  %%CITATION = PHLTA,B644,153;%%


\bibitem{Fritzsch:1979zq}
  H.~Fritzsch,
  %``Quark Masses And Flavor Mixing,''
  Nucl.\ Phys.\  B {\bf 155} (1979) 189;
  %%CITATION = NUPHA,B155,189;%%
%\cite{Fritzsch:1999ee}


\bibitem{Fritzsch:1999ee}
  H.~Fritzsch and Z.~z.~Xing,
  %``Mass and flavor mixing schemes of quarks and leptons,''
  Prog.\ Part.\ Nucl.\ Phys.\  {\bf 45} (2000) 1
  [arXiv:hep-ph/9912358].
  %%CITATION = PPNPD,45,1;%%

\bibitem{textures}
%\cite{Leontaris:2009pi}
%\bibitem{Leontaris:2009pi}
  G.~K.~Leontaris and N.~D.~Vlachos,
  %``D-brane Inspired Fermion Mass Textures,''
  arXiv:0909.4701 [Unknown];
  %%CITATION = ARXIV:0909.4701;%%
%\cite{Dev:2009he}
%\bibitem{Dev:2009he}
  S.~Dev, S.~Verma and S.~Gupta,
  %``Phenomenological Analysis of Hybrid Textures of Neutrinos,''
  arXiv:0909.3182 [Unknown];
  %%CITATION = ARXIV:0909.3182;%%
%\cite{Adhikary:2009kz}
%\bibitem{Adhikary:2009kz}
  B.~Adhikary, A.~Ghosal and P.~Roy,
  %``'Mu-Tau' symmetry, tribimaximal mixing and four zero neutrino Yukawa
  %textures,''
  arXiv:0908.2686 [hep-ph];
  %%CITATION = ARXIV:0908.2686;%%
%\cite{Goswami:2009bd}
%\bibitem{Goswami:2009bd}
  S.~Goswami, S.~Khan and W.~Rodejohann,
  %``Minimal Textures in Seesaw Mass Matrices and their low and high Energy
  %Phenomenology,''
  Phys.\ Lett.\  B {\bf 680} (2009) 255
  [arXiv:0905.2739 [hep-ph]];
  %%CITATION = PHLTA,B680,255;%%
%\cite{Goswami:2008uv}
%\bibitem{Goswami:2008uv}
  S.~Goswami, S.~Khan and A.~Watanabe,
  %``Hybrid textures in minimal seesaw mass matrices,''
  arXiv:0811.4744 [hep-ph];
  %%CITATION = ARXIV:0811.4744;%%
%\cite{Choubey:2008tb}
%\bibitem{Choubey:2008tb}
  S.~Choubey, W.~Rodejohann and P.~Roy,
  %``Phenomenological consequences of four zero neutrino Yukawa textures,''
  Nucl.\ Phys.\  B {\bf 808} (2009) 272
  [Erratum-ibid.\  {\bf 818} (2009) 136]
  [arXiv:0807.4289 [hep-ph]];
  %%CITATION = NUPHA,B808,272;%%
%\cite{Branco:2007nb}
%\bibitem{Branco:2007nb}
  G.~C.~Branco, D.~Emmanuel-Costa, M.~N.~Rebelo and P.~Roy,
  %``Four Zero Neutrino Yukawa Textures in the Minimal Seesaw Framework,''
  Phys.\ Rev.\  D {\bf 77} (2008) 053011
  [arXiv:0712.0774 [hep-ph]];
  %%CITATION = PHRVA,D77,053011;%%
%\cite{Alhendi:2007iu}
%\bibitem{Alhendi:2007iu}
  H.~A.~Alhendi, E.~I.~Lashin and A.~A.~Mudlej,
  %``Textures with two traceless submatrices of the neutrino mass matrix,''
  Phys.\ Rev.\  D {\bf 77} (2008) 013009
  [arXiv:0708.2007 [hep-ph]];
  %%CITATION = PHRVA,D77,013009;%%
%\cite{Kaneko:2007ea}
%\bibitem{Kaneko:2007ea}
  S.~Kaneko, H.~Sawanaka, T.~Shingai, M.~Tanimoto and K.~Yoshioka,
  %``New approach to texture-zeros with S(3) symmetry: Flavor symmetry and
  %vacuum aligned mass textures,''
  arXiv:hep-ph/0703250;
  %%CITATION = HEP-PH/0703250;%%
%\cite{Branco:2006wv}
%\bibitem{Branco:2006wv}
  G.~C.~Branco, M.~N.~Rebelo and J.~I.~Silva-Marcos,
  %``Yukawa Textures, New Physics and Nondecoupling,''
  Phys.\ Rev.\  D {\bf 76} (2007) 033008
  [arXiv:hep-ph/0612252];
  %%CITATION = PHRVA,D76,033008;%%
%\cite{Lam:2006wm}
%\bibitem{Lam:2006wm}
  C.~S.~Lam,
  %``Mass Independent Textures and Symmetry,''
  Phys.\ Rev.\  D {\bf 74} (2006) 113004
  [arXiv:hep-ph/0611017];
  %%CITATION = PHRVA,D74,113004;%%
%\cite{Kaneko:2006wi}
%\bibitem{Kaneko:2006wi}
  S.~Kaneko, H.~Sawanaka, T.~Shingai, M.~Tanimoto and K.~Yoshioka,
  %``Flavor Symmetry and Vacuum Aligned Mass Textures,''
  Prog.\ Theor.\ Phys.\  {\bf 117} (2007) 161
  [arXiv:hep-ph/0609220];
  %%CITATION = PTPKA,117,161;%%
%\cite{Fuki:2006xw}
%\bibitem{Fuki:2006xw}
  K.~Fuki and M.~Yasue,
  %``Two categories of approximately mu - tau symmetric neutrino mass
  %textures,''
  Nucl.\ Phys.\  B {\bf 783} (2007) 31
  [arXiv:hep-ph/0608042];
  %%CITATION = NUPHA,B783,31;%%
%\cite{Haba:2005ds}
%\bibitem{Haba:2005ds}
  N.~Haba and K.~Yoshioka,
  %``Discrete flavor symmetry, dynamical mass textures, and grand
  %unification,''
  Nucl.\ Phys.\  B {\bf 739} (2006) 254
  [arXiv:hep-ph/0511108];
  %%CITATION = NUPHA,B739,254;%%
%\cite{Kim:2004ki}
%\bibitem{Kim:2004ki}
  H.~D.~Kim, S.~Raby and L.~Schradin,
  %``Quark mass textures and sin(2beta),''
  Phys.\ Rev.\  D {\bf 69} (2004) 092002
  [arXiv:hep-ph/0401169];
  %%CITATION = PHRVA,D69,092002;%%
%\cite{Jack:2003pb}
%\bibitem{Jack:2003pb}
  I.~Jack, D.~R.~T.~Jones and R.~Wild,
  %``Yukawa textures and the mu-term,''
  Phys.\ Lett.\  B {\bf 580} (2004) 72
  [arXiv:hep-ph/0309165];
  %%CITATION = PHLTA,B580,72;%%
%\cite{Jack:2003qg}
%\bibitem{Jack:2003qg}
  I.~Jack and D.~R.~T.~Jones,
  %``Yukawa textures and anomaly mediated supersymmetry breaking,''
  Nucl.\ Phys.\  B {\bf 662} (2003) 63
  [arXiv:hep-ph/0301163];
  %%CITATION = NUPHA,B662,63;%%
%\cite{Caravaglios:2002br}
%\bibitem{Caravaglios:2002br}
  F.~Caravaglios, P.~Roudeau and A.~Stocchi,
  %``Precision test of quark mass textures: A Model independent approach,''
  Nucl.\ Phys.\  B {\bf 633} (2002) 193
  [arXiv:hep-ph/0202055];
  %%CITATION = NUPHA,B633,193;%%
%\cite{Everett:2000up}
%\bibitem{Everett:2000up}
  L.~L.~Everett, G.~L.~Kane and S.~F.~King,
  %``D branes and textures,''
  JHEP {\bf 0008} (2000) 012
  [arXiv:hep-ph/0005204];
  %%CITATION = JHEPA,0008,012;%%
%\cite{Berezhiani:2000cg}
%\bibitem{Berezhiani:2000cg}
  Z.~Berezhiani and A.~Rossi,
  %``Predictive grand unified textures for quark and neutrino masses and
  %mixings,''
  Nucl.\ Phys.\  B {\bf 594} (2001) 113
  [arXiv:hep-ph/0003084];
  %%CITATION = NUPHA,B594,113;%%
%\cite{Kuo:1999dt}
%\bibitem{Kuo:1999dt}
  T.~K.~Kuo, S.~W.~Mansour and G.~H.~Wu,
  %``Triangular textures for quark mass matrices,''
  Phys.\ Rev.\  D {\bf 60} (1999) 093004
  [arXiv:hep-ph/9907314];
  %%CITATION = PHRVA,D60,093004;%%
%\cite{Falcone:1998us}
%\bibitem{Falcone:1998us}
  D.~Falcone and F.~Tramontano,
  %``Relation between quark masses and weak mixings,''
  Phys.\ Rev.\  D {\bf 59} (1999) 017302
  [arXiv:hep-ph/9806496].
  %%CITATION = PHRVA,D59,017302;%%



%\cite{Roberts:2001zy}
\bibitem{Roberts:2001zy}
  R.~G.~Roberts, A.~Romanino, G.~G.~Ross and L.~Velasco-Sevilla,
  %``Precision test of a Fermion mass texture,''
  Nucl.\ Phys.\  B {\bf 615} (2001) 358
  [arXiv:hep-ph/0104088];
  %%CITATION = NUPHA,B615,358;%%
%\cite{Ramond:1993kv}
%\bibitem{Ramond:1993kv}
  P.~Ramond, R.~G.~Roberts and G.~G.~Ross,
  %``Stitching the Yukawa quilt,''
  Nucl.\ Phys.\  B {\bf 406} (1993) 19
  [arXiv:hep-ph/9303320].
  %%CITATION = NUPHA,B406,19;%%


%\cite{Chiu:2000gw}
\bibitem{Chiu:2000gw}
  S.~H.~Chiu, T.~K.~Kuo and G.~H.~Wu,
  %``Hermitian quark mass matrices with four texture zeros,''%\cite{Fritzsch:1995nx}
  Phys.\ Rev.\  D {\bf 62} (2000) 053014 [arXiv:hep-ph/0003224].



%\cite{Fritzsch:1999rb}
\bibitem{Fritzsch:1999rb}
  H.~Fritzsch and Z.~z.~Xing,
  %``The light quark sector, CP violation, and the unitarity triangle,''
  Nucl.\ Phys.\  B {\bf 556} (1999) 49
  [arXiv:hep-ph/9904286];
  %%CITATION = NUPHA,B556,49;%%
%\cite{Fritzsch:2002ga}
%\bibitem{Fritzsch:2002ga}
  H.~Fritzsch and Z.~z.~Xing,
  %``Four-zero texture of Hermitian quark mass matrices and current experimental
  %tests,''
  Phys.\ Lett.\  B {\bf 555} (2003) 63
  [arXiv:hep-ph/0212195].
  %%CITATION = PHLTA,B555,63;%%


\bibitem{sumrule}
S.~F.~King,
%``Predicting neutrino parameters from SO(3) family symmetry and quark-lepton
%unification,''
JHEP {\bf 0508} (2005) 105;
%[arXiv:hep-ph/0506297];
I.~Masina,
  %``A maximal atmospheric mixing from a maximal CP violating phase,''
  Phys.\ Lett.\  B {\bf 633} (2006) 134;
%[arXiv:hep-ph/0508031];
  %%CITATION = PHLTA,B633,134;%%
S.~Antusch and S.~F.~King,
%   ``Charged lepton corrections to neutrino mixing angles and CP phases
  %revisited,''
  Phys.\ Lett.\ B {\bf 631} (2005) 42.
%[arXiv:hep-ph/0508044];
  %%CITATION = HEP-PH 0508044;%%

%\cite{Antusch:2008yc}
\bibitem{Antusch:2008yc}
  S.~Antusch, S.~F.~King and M.~Malinsky,
  %``Perturbative Estimates of Lepton Mixing Angles in Unified Models,''
  arXiv:0810.3863 [hep-ph];
  %%CITATION = ARXIV:0810.3863;%%
%\cite{Boudjemaa:2008jf}
%\bibitem{Boudjemaa:2008jf}
  S.~Boudjemaa and S.~F.~King,
  %``Deviations from Tri-bimaximal Mixing: Charged Lepton Corrections and
  %Renormalization Group Running,''
  arXiv:0808.2782 [hep-ph];
  %%CITATION = ARXIV:0808.2782;%%
%\cite{Antusch:2007vw}
%\bibitem{Antusch:2007vw}
  S.~Antusch, S.~F.~King and M.~Malinsky,
  %``Third Family Corrections to Quark and Lepton Mixing in SUSY Models with
  %non-Abelian Family Symmetry,''
  JHEP {\bf 0805} (2008) 066
  [arXiv:0712.3759 [hep-ph]];
  %%CITATION = JHEPA,0805,066;%%
%\cite{Antusch:2007ib}
%\bibitem{Antusch:2007ib}
  S.~Antusch, S.~F.~King and M.~Malinsky,
  %``Third Family Corrections to Tri-bimaximal Lepton Mixing and a New Sum
  %Rule,''
  Phys.\ Lett.\  B {\bf 671} (2009) 263
  [arXiv:0711.4727 [hep-ph]];
  %%CITATION = PHLTA,B671,263;%%
%\cite{King:2007pr}
%\bibitem{King:2007pr}
  S.~F.~King,
  %``Parametrizing the lepton mixing matrix in terms of deviations from
  %tri-bimaximal mixing,''
  Phys.\ Lett.\  B {\bf 659} (2008) 244
  [arXiv:0710.0530 [hep-ph]];
  %%CITATION = PHLTA,B659,244;%%
%\cite{Antusch:2007rk}
%\bibitem{Antusch:2007rk}
  S.~Antusch, P.~Huber, S.~F.~King and T.~Schwetz,
  %``Neutrino mixing sum rules and oscillation experiments,''
  JHEP {\bf 0704} (2007) 060
  [arXiv:hep-ph/0702286];
  %%CITATION = JHEPA,0704,060;%%
%\cite{Antusch:2005kw}
%\bibitem{Antusch:2005kw}
  S.~Antusch and S.~F.~King,
  %``Charged lepton corrections to neutrino mixing angles and CP phases
  %revisited,''
  Phys.\ Lett.\  B {\bf 631} (2005) 42
  [arXiv:hep-ph/0508044].
  %%CITATION = PHLTA,B631,42;%%

\bibitem{CKM}
CKMfitter Group (J. Charles et al.), Eur. Phys. J. C41 (2005) 1, hep-ph/0406184, updated
results and plots: http://ckmfitter.in2p3.fr; M. Bona et al. (UTfit Collaboration),
hep-ph/0701204.


\bibitem{Xing}
  H.~Fritzsch and Z.~Z.~Xing,
  %``Flavor symmetries and the description of flavor mixing,''
  Phys.\ Lett.\  B {\bf 413} (1997) 396
  [arXiv:hep-ph/9707215];
  %%CITATION = PHLTA,B413,396;%%
  H.~Fritzsch and Z.~z.~Xing,
  %``On the parametrization of flavor mixing in the standard model,''
  Phys.\ Rev.\  D {\bf 57} (1998) 594
  [arXiv:hep-ph/9708366].
  %%CITATION = PHRVA,D57,594;%%

\bibitem{Scott}
  P.~F.~Harrison, D.~R.~J.~Roythorne and W.~G.~Scott,
  %``Plaquette Invariants and the Flavour Symmetric Description of Quark and
  %Neutrino Mixings,''
  Phys.\ Lett.\  B {\bf 657} (2007) 210
  [arXiv:0709.1439 [hep-ph]];
  %%CITATION = PHLTA,B657,210;%%
  P.~F.~Harrison, D.~R.~J.~Roythorne and W.~G.~Scott,
  %``Flavour Permutation Symmetry and Fermion Mixing,''
  arXiv:0805.3440 [hep-ph];
  %%CITATION = ARXIV:0805.3440;%%
  P.~F.~Harrison, D.~R.~J.~Roythorne and W.~G.~Scott,
  %``Is the Unitarity Triangle Right?,''
  arXiv:0904.3014 [hep-ph];
  %%CITATION = ARXIV:0904.3014;%%
  P.~F.~Harrison, S.~Dallison and W.~G.~Scott,
  %``The Matrix of Unitarity Triangle Angles for Quarks,''
  Phys.\ Lett.\  B {\bf 680} (2009) 328
  [arXiv:0904.3077 [hep-ph]];
  %%CITATION = PHLTA,B680,328;%%
  P.~F.~Harrison and W.~G.~Scott,
  %``A Flavour-Symmetric Perspective on Neutrino Mixing,''
  arXiv:0906.2732 [hep-ph].
  %%CITATION = ARXIV:0906.2732;%%

\bibitem{Toharia}
  G.~Couture, C.~Hamzaoui, S.~S.~Y.~Lu and M.~Toharia,
  %``Patterns in the Fermion Mixing Matrix, a bottom-up approach,''
  arXiv:0910.3132 [Unknown].
  %%CITATION = ARXIV:0910.3132;%%



\bibitem{PDG}
  C. Amsler {\it et al.}  [Particle Data Group],
  Physics Letters B667 (2008) 1.

%\cite{Xing:2009eg}
\bibitem{Xing:2009eg}
  Z.~z.~Xing,
  %``Right Unitarity Triangles, Stable CP-violating Phases and Approximate
  %Quark-Lepton Complementarity,''
  Phys.\ Lett.\  B {\bf 679} (2009) 111
  [arXiv:0904.3172 [hep-ph]].
  %%CITATION = PHLTA,B679,111;%%

%\cite{King:2002nf}
\bibitem{King:2002nf}
  S.~F.~King,
  %``Constructing the large mixing angle MNS matrix in see-saw models with
  %right-handed neutrino dominance,''
  JHEP {\bf 0209} (2002) 011
  [arXiv:hep-ph/0204360].
  %%CITATION = JHEPA,0209,011;%%


%\cite{Barr:1990td}
\bibitem{Barr:1990td}
  S.~M.~Barr,
  %``A Predictive hierarchical mode of quark and lepton masses,''
  Phys.\ Rev.\  D {\bf 42} (1990) 3150;
  %%CITATION = PHRVA,D42,3150;%%
%\cite{Sogami:1992av}
%\bibitem{Sogami:1992av}
  I.~S.~Sogami and T.~Shinohara,
  %``Universal seesaw mechanisms for quark lepton mass spectrum,''
  Phys.\ Rev.\  D {\bf 47} (1993) 2905.
  %%CITATION = PHRVA,D47,2905;%%
%\cite{Ibanez:1994ig}
%\bibitem{Ibanez:1994ig}
  L.~E.~Ibanez and G.~G.~Ross,
  %``Fermion masses and mixing angles from gauge symmetries,''
  Phys.\ Lett.\  B {\bf 332} (1994) 100
  [arXiv:hep-ph/9403338];
  %%CITATION = PHLTA,B332,100;%%
%\cite{Jain:1994hd}
%\bibitem{Jain:1994hd}
  V.~Jain and R.~Shrock,
  %``Models of fermion mass matrices based on a flavor dependent and generation
  %dependent U(1) gauge symmetry,''
  Phys.\ Lett.\  B {\bf 352} (1995) 83
  [arXiv:hep-ph/9412367];
  %%CITATION = PHLTA,B352,83;%%
%\cite{Pomarol:1995xc}
%\bibitem{Pomarol:1995xc}
  A.~Pomarol and D.~Tommasini,
  %``Horizontal symmetries for the supersymmetric flavor problem,''
  Nucl.\ Phys.\  B {\bf 466} (1996) 3
  [arXiv:hep-ph/9507462];
  %%CITATION = NUPHA,B466,3;%%
%\cite{Mondragon:1998gy}
%\bibitem{Mondragon:1998gy}
  A.~Mondragon and E.~Rodriguez-Jauregui,
  %``The breaking of the flavour permutational symmetry: Mass textures and the
  %CKM matrix,''
  Phys.\ Rev.\  D {\bf 59} (1999) 093009
  [arXiv:hep-ph/9807214];
  %%CITATION = PHRVA,D59,093009;%%
%\cite{Barbieri:1999pe}
%\bibitem{Barbieri:1999pe}
  R.~Barbieri, P.~Creminelli and A.~Romanino,
  %``Neutrino mixings from a U(2) flavour symmetry,''
  Nucl.\ Phys.\  B {\bf 559} (1999) 17
  [arXiv:hep-ph/9903460];
  %%CITATION = NUPHA,B559,17;%%
%\cite{Barbieri:1998em}
%\bibitem{Barbieri:1998em}
  R.~Barbieri, L.~Giusti, L.~J.~Hall and A.~Romanino,
  %``Fermion masses and symmetry breaking of a U(2) flavour symmetry,''
  Nucl.\ Phys.\  B {\bf 550} (1999) 32
  [arXiv:hep-ph/9812239];
  %%CITATION = NUPHA,B550,32;%%
%\cite{Babu:2004tn}
%\bibitem{Babu:2004tn}
  K.~S.~Babu and J.~Kubo,
  %``Dihedral families of quarks, leptons and Higgses,''
  Phys.\ Rev.\  D {\bf 71} (2005) 056006
  [arXiv:hep-ph/0411226];
  %%CITATION = PHRVA,D71,056006;%%
%\cite{Haba:2005ds}
%\bibitem{Haba:2005ds}
  N.~Haba and K.~Yoshioka,
  %``Discrete flavor symmetry, dynamical mass textures, and grand
  %unification,''
  Nucl.\ Phys.\  B {\bf 739} (2006) 254
  [arXiv:hep-ph/0511108];
  %%CITATION = NUPHA,B739,254;%%
%\cite{Masina:2006pe}
%\bibitem{Masina:2006pe}
  I.~Masina and C.~A.~Savoy,
  %``Up quark masses from down quark masses,''
  Phys.\ Lett.\  B {\bf 642} (2006) 472
  [arXiv:hep-ph/0606097];
  %%CITATION = PHLTA,B642,472;%%
%\cite{Bazzocchi:2007na}
%\bibitem{Bazzocchi:2007na}
  F.~Bazzocchi, S.~Kaneko and S.~Morisi,
  %``A SUSY A4 model for fermion masses and mixings,''
  JHEP {\bf 0803} (2008) 063
  [arXiv:0707.3032 [hep-ph]];
  %%CITATION = JHEPA,0803,063;%%
%\cite{Babu:2009nn}
%\bibitem{Babu:2009nn}
  K.~S.~Babu and Y.~Meng,
  %``Flavor Violation in Supersymmetric Q_6 Model,''
  Phys.\ Rev.\  D {\bf 80} (2009) 075003
  [arXiv:0907.4231 [hep-ph]];
  %%CITATION = PHRVA,D80,075003;%%
%\cite{Ishimori:2009ns}
%\bibitem{Ishimori:2009ns}
  H.~Ishimori, Y.~Shimizu and M.~Tanimoto,
  %``S4 Flavor Model of Quarks and Leptons,''
  Prog.\ Theor.\ Phys.\ Suppl.\  {\bf 180} (2010) 61
  [arXiv:0904.2450 [hep-ph]];
  %%CITATION = PTPSA,180,61;%%
%\cite{Morisi:2010rk}
%\bibitem{Morisi:2010rk}
  S.~Morisi and E.~Peinado,
  %``An S4 model for quarks and leptons with maximal atmospheric angle,''
  arXiv:1001.2265 [hep-ph].
  %%CITATION = ARXIV:1001.2265;%%


\bibitem{GST}
  R. Gatto, G. Sartori and M. Tonin, Phys. Lett. B {\bf 28} (1968) 128.
%\cite{Fritzsch:1979zq}

%\cite{Xing:2007fb}
\bibitem{Xing:2007fb}
  Z.~z.~Xing, H.~Zhang and S.~Zhou,
  %``Updated Values of Running Quark and Lepton Masses,''
  arXiv:0712.1419 [hep-ph].
  %%CITATION = ARXIV:0712.1419;%%


\bibitem{GJ}
  H.~Georgi and C.~Jarlskog,
  %``A New Lepton - Quark Mass Relation In A Unified Theory,''
  Phys.\ Lett.\  B {\bf 86} (1979) 297.
  %%CITATION = PHLTA,B86,297;%%

\bibitem{Schwetz:2008er}
  T.~Schwetz, M.~Tortola and J.~W.~F.~Valle,
  %``Three-flavour neutrino oscillation update,''
  New J.\ Phys.\  {\bf 10} (2008) 113011
  [arXiv:0808.2016 [hep-ph]].
  %%CITATION = NJOPF,10,113011;%%


\bibitem{Antusch:2009gu}
  S.~Antusch and M.~Spinrath,
  %``New GUT predictions for quark and lepton mass ratios confronted with
  %phenomenology,''
  arXiv:0902.4644 [hep-ph].
  %%CITATION = ARXIV:0902.4644;%%

\bibitem{RGE}
  S.~Antusch, J.~Kersten, M.~Lindner and M.~Ratz,
  %``Running neutrino masses, mixings and CP phases: Analytical results and
  %phenomenological consequences,''
  Nucl.\ Phys.\  B {\bf 674} (2003) 401;
%  [arXiv:hep-ph/0305273].
S.~Antusch, J.~Kersten, M.~Lindner, M.~Ratz and M.~A.~Schmidt,
  %``Running neutrino mass parameters in see-saw scenarios,''
  JHEP {\bf 0503} (2005) 024
  [arXiv:hep-ph/0501272].
  %%CITATION = JHEPA,0503,024;%%

%\cite{King:2009qt}
\bibitem{King:2009qt}
  S.~F.~King,
  %``Tri-bimaximal Neutrino Mixing and $\theta_{13}$,''
  arXiv:0903.3199 [hep-ph].
  %%CITATION = ARXIV:0903.3199;%%


\end{thebibliography}
\end{document}